\documentclass{amsart}
\usepackage{amssymb}
\usepackage{amsmath}
\usepackage{amsfonts}

\newtheorem{theorem}{Theorem}
\theoremstyle{plain}

\newtheorem{lemma}{Lemma}

\numberwithin{equation}{section}
\input{tcilatex}

\begin{document}
\title[Inverse scattering transform for N-wave interaction problem ]{Inverse
scattering transform for N-wave interaction problem with a dispersive term
in two spatial dimensions}
\author{Mansur I Ismailov$^{\ast ),\ast \ast )}$}
\address{$^{\ast )}$Department of Mathematics, Gebze Technical University,
Gebze-Kocaeli \ 41400, Turkey \\
$^{\ast \ast )}$Institute of Mathematics and Mechanics, Azerbaijan National
Academy of Science, 1141 Baku, Azerbaijan}
\email{mismailov@gtu.edu.tr }
\date{20, August, 2020}
\dedicatory{}

\begin{abstract}
In this work, we introduce a dispersive $N$-wave interaction problem ($%
N=2n,n\in 
\mathbb{N}
$) involving $n$ velocities in two spatial dimensions and one temporal
dimension. Exact solutions of the problem are exhibited. This is a
generalization of the $N$-wave interaction problem and matrix
Davey-Stewartson equation with 2+1 dimensions that examines the Benney-type
model of interactions between short and long waves. Accordingly, associated
with the solutions of two dimensional analog of the Manakov system, a
Gelfand-Levitan-Marchenko (GLM)-type, or so-called inversion-like, equation
is constructed. It is shown that the presence of the degenerate kernel reads
exact soliton-like solutions of the dispersive $N$-wave interaction
problem.We also mention the unique solution of the Cauchy problem on an
arbitrary time interval for small initial data.
\end{abstract}

\maketitle

\section{Introduction}

The inverse scattering transform (IST) for nonlinear evolution equations
with 2+1, i.e., two spatial and one temporal dimensions has been started
with the papers by Zakharov and Shabat [18, 19]. For the general case of
evolutional partial differential equations in 2+1 dimensions the IST
requires a novel approach, namely either a nonlocal Riemann--Hilbert (RH)
[3] or a $\partial $-bar formalism [1, 4], however for certain nonlinear
two-dimensional equations, the classical approach of the IST via the GLM
equation is still applicable [2, 11, 12]. The IST can be employed to the
initial value problem for a variety of physically significant equations
which are related to the inverse scattering problem for first order systems
of partial differential equations. Concrete results with a wide class of the
exact solutions for various forms of Davey--Stewartson,
Kadomtsev--Petviashvili equations and the \textit{N}-wave interaction in 2 +
1 dimension, has been obtained in [7, 13] on the basis of the analysis of
GLM type integral equations, in [5, 17] on the basis of the nonlocal
Riemann--Hilbert problem and in [6] via the $\partial $-bar method.

This paper considers the two-dimensional spatial dispersive $2n$-wave
interaction problem with $n$ velocities which is generalized the $N-$wave
interaction problem of [7] and two component Davey--Stewartson\ equation of
[13]. This nonlinear equation admits a Lax-type representation. Therefore we
use the IST via the GLM equation for its integration. As the inverse problem
we set the two dimensional inverse-scattering problem for the following
Manakov system, studied in detail in [8]:

\begin{equation}
\frac{\partial }{\partial y}\mathbf{\psi }-\mathbf{\sigma }\frac{\partial }{%
\partial x}\mathbf{\psi }+\mathbf{Q}(x,y)\mathbf{\psi }=\mathbf{0}, 
\tag{1.1}
\end{equation}%
where $\mathbf{\sigma }=\left( 
\begin{array}{cc}
\mathbf{I}_{n} & \mathbf{0}_{n\times 1} \\ 
\mathbf{0}_{1\times n} & -1%
\end{array}%
\right) $ is constant $n\times n$ diagonal matrix with the identity matrix $%
\mathbf{I}_{n}$ of the order $n$ and $n\times 1$ column $\mathbf{0}_{n\times
1}$ and $1\times n$ row $\mathbf{0}_{1\times n}$ zero vectors; $\mathbf{Q}%
=\left( 
\begin{array}{cc}
\mathbf{0}_{n} & \mathbf{q}_{12} \\ 
\mathbf{q}_{21} & 0%
\end{array}%
\right) $ is an off-diagonal matrix with the zero matrix entry $\mathbf{0}%
_{n}$ of the order $n$ and the $n\times 1$ column $\mathbf{q}_{12}$ and $%
1\times n$ row $\mathbf{q}_{21}$vector functions.

The case $n=1$ makes this system two component nonstationary Dirac system
[14] (and also two dimesional analogue of Zakharov-Shabat (ZS) or AKNS
system [15]). In the case $n=2$ this system is two dimesional analogue of
Manakov system [10] and for arbitrary positive integer $n$ is the two
dimesional analogue of Dirac-type sysytem [16]. Inverse scattering theory
for the system (1.1) is satisfactoryly investigated in [14] for the case $%
n=1 $ and in [8] for arbitrary $n>1$. This paper use this inverse problem to
solve dispersive $2n$-wave interaction problem with $n$ velocities satisfied
by additionally time dependent $\mathbf{Q}(x,y;t)$, which is generalized $2n$
-wave interaction with $n$ velocities\ and $n\times n$ matrix
Davey-Stewartson\ equation in two spatial and one temporal dimensions.

This article is organized as follows: In Section 2, the dispersive $2n$-wave
interaction problem with $n$ velocities and its Lax representation is
introduced. It was clear that its spectral problem is the problem two
dimensional analogue of Manakov system (1.1) that the inverse scattering
problem is studied in detail in [8]. Section 3, deals with the inverse
problem associated with the linear equation (1.1) and the corresponding
multidimensional GLM equation with explicitly solvable degenerate kernel
case. The aim of the Section 4 is to apply the results of [8] to the
integration of the dispersive 2n-wave interaction problem with n velocities
by using the IST method: The Cauchy problem is investigated the exact
soliton-like solutions are derived.

\section{\textbf{Dispersive }$\mathbf{2n}$\textbf{-wave interaction problem
with }$\mathbf{n}$\textbf{\ velocities and its Lax representation}}

Consider the system of $N-$wave equations with $N=2n$ in the following form: 
\begin{eqnarray}
\frac{\partial }{\partial t}q_{k}+\alpha _{k}\frac{\partial }{\partial x}%
q_{k}+\beta _{k}\frac{\partial }{\partial y}q_{k}-i\gamma \frac{\partial ^{2}%
}{\partial x\partial y}q_{k} &=&\dsum\limits_{m=1}^{n}p_{km}q_{m}-pq_{k}, 
\notag \\
&&  \TCItag{2.1} \\
\frac{\partial }{\partial t}q_{n+k}+\alpha _{k}\frac{\partial }{\partial x}%
q_{n+k}+\beta _{k}\frac{\partial }{\partial y}q_{n+k}+i\gamma \frac{\partial
^{2}}{\partial x\partial y}q_{n+k}
&=&-\dsum\limits_{m=1}^{n}p_{mk}q_{n+m}+pq_{n+k},  \notag \\
k &=&1,...,n  \notag
\end{eqnarray}%
where $\alpha _{k},$ $\beta _{k}$ and $\gamma $ are real numbers with $\beta
_{k}-\alpha _{k}\neq \beta _{j}-\alpha _{j}$ and $\beta _{k}+\alpha
_{k}=\beta _{j}+\alpha _{j}$ when $k\neq j$. The functions $p_{km}$ and $p$
are solutions of the equations 
\begin{eqnarray}
\frac{\partial }{\partial \xi }p &=&-i\gamma \dsum\limits_{m=1}^{n}\frac{%
\partial }{\partial x}(q_{m}q_{n+m}),\text{ }\frac{\partial }{\partial \eta }%
p_{kk}=-i\gamma \frac{\partial }{\partial x}(q_{k}q_{n+k}),\text{ }k=1,...,n,
\notag \\
\frac{\partial }{\partial \eta }p_{km} &=&\left( \beta _{m}-\beta
_{k}\right) q_{k}q_{n+m}-i\gamma \frac{\partial }{\partial x}(q_{k}q_{n+m}),%
\text{ }m,k=1,...,n;\text{ }m\neq k.  \TCItag{2.2}
\end{eqnarray}%
where $\frac{\partial }{\partial \xi }=\frac{\partial }{\partial y}+\frac{%
\partial }{\partial x},\frac{\partial }{\partial \eta }=\frac{\partial }{%
\partial y}-\frac{\partial }{\partial x}.$

This system (2.1) is the 2+1 dimensional $N$-wave interaction problem with
the dispersive term $i\gamma \frac{\partial ^{2}}{\partial x\partial y}q_{k}$
and also with the quasi-potential (2.2). The aim of this paper is the
integrability of this system by using the suitable method of inverse
scattering transform. The case when the terms $\alpha _{k}\frac{\partial }{%
\partial x}q_{k}+\beta _{k}\frac{\partial }{\partial y}q_{k}$ absence the
equation (2.1) becomes to 2+1 dimensional nonlinear Schrodinger equation and
the IST method for its integration is realized in [13] for $n=1$ by using
the inverse scattering problem (ISP) for two component nonstationary Dirac
equation, [14]. The undispresive system (2.1) when $\gamma =0$, can be
integrate by using IST method in [7, 9] which the integrate ISP is matrix
nonstationary Dirac system of $n+1$ components with $n>1$, [8].

Generally, in physical systems, waves with different length scales appear.
They are examine the interactions between waves for certain model partial
differential equations. In Benney model [20], the equations (2.1), (2.2)
examines the interactions between short and long waves, where $p$ is the
long wave profile and $q$ is, to leading order, the short wave envelope. The
constants $\alpha _{k}$ and $\beta _{k}$ are the group velocities of the
short waves, $\gamma $ is due to the linear dispersion.

Let us denote $\mathbf{q}_{12}=\func{col}\left\{ q_{1},...,q_{n}\right\} $, $%
\mathbf{q}_{21}=\func{col}\left\{ q_{n+1},...,q_{2n}\right\} ^{T},$ $\mathbf{%
P}=(p_{km})_{k,m=1}^{n}$. Then, the equation (2.1) and (2.2) are reduced to
the matrix system 
\begin{eqnarray}
\partial _{t}\mathbf{q}_{12}-\mathbf{B}\partial _{\xi }\mathbf{q}%
_{12}+b\partial _{\eta }\mathbf{q}_{12}-i\gamma (\mathbf{q}_{12})_{xy} &=&%
\mathbf{Pq}_{12}-p\mathbf{q}_{12},  \notag \\
&&  \TCItag{2.3} \\
\partial _{t}\mathbf{q}_{21}-\partial _{\xi }\mathbf{q}_{21}\mathbf{B}%
+b\partial _{\eta }\mathbf{q}_{21}+i\gamma (\mathbf{q}_{21})_{xy} &=&p%
\mathbf{q}_{21}-\mathbf{q}_{21}\mathbf{P},  \notag
\end{eqnarray}%
and 
\begin{eqnarray}
\partial _{\eta }\mathbf{P} &=&\left[ \mathbf{q}_{12}\mathbf{q}_{21}\mathbf{%
,B}\right] -i\gamma (\mathbf{q}_{12}\mathbf{q}_{21})_{x},  \notag \\
&&  \TCItag{2.4} \\
\partial _{\xi }p &=&-i\gamma (\mathbf{q}_{21}\mathbf{q}_{12})_{x},  \notag
\end{eqnarray}%
respectively, where $\partial _{t}=\frac{\partial }{\partial t},\partial
_{\xi }=\frac{\partial }{\partial y}+\frac{\partial }{\partial x},\partial
_{\eta }=\frac{\partial }{\partial y}-\frac{\partial }{\partial x},$ $%
\mathbf{B}=diag(b_{1},\ldots ,b_{n})$ with $b_{k}=\frac{\beta _{k}-\alpha
_{k}}{2}$ and $b=-\frac{\beta _{k}+\alpha _{k}}{2}$.

\bigskip Let $\mathbf{M}$ and $\mathbf{A}$ be first order matrix operators:%
\begin{equation*}
\mathbf{M}=\mathbf{\sigma }\frac{\partial }{\partial x}+\mathbf{Q},\text{ \ }%
\mathbf{A}=\mathbf{\delta }\frac{\partial }{\partial x}+i\gamma \mathbf{I}%
_{n}\frac{\partial ^{2}}{\partial x^{2}}+\mathbf{\Gamma }.
\end{equation*}

Then the equation (2.3), (2.4) admits the following Lax representation: 
\begin{equation}
\left[ \frac{\partial }{\partial y}-\mathbf{M},\frac{\partial }{\partial t}-%
\mathbf{A}\right] =0.  \tag{2.5}
\end{equation}

Here $\mathbf{\delta }$ and $\mathbf{\Gamma }$\textbf{\ }are $n+1$-th\ ($%
n\geq 2$) order square matrices. The matrix $\mathbf{\delta }$ be a real and
diagonal: $\mathbf{\delta }=\left( 
\begin{array}{cc}
2\mathbf{B} & \mathbf{0}_{n\times 1} \\ 
\mathbf{0}_{1\times n} & 2b%
\end{array}%
\right) ,$ where $\mathbf{B}=diag(b_{1},\ldots ,b_{n})$ with $b_{k}\neq
b_{j}\neq b$ when $k\neq j$ and $\mathbf{\Gamma }$ is the following form $%
\mathbf{\Gamma }=\left( 
\begin{array}{cc}
\mathbf{P} & \left( \mathbf{B}-b\mathbf{I}_{n}\right) \mathbf{q}_{12} \\ 
\mathbf{q}_{21}\left( \mathbf{B}-b\mathbf{I}_{n}\right) & p%
\end{array}%
\right) $ that obey the relation $\left[ \mathbf{\sigma ,\Gamma }\right] =%
\left[ \mathbf{\delta ,Q}\right] +2i\gamma \mathbf{Q}_{x}$.

Let us denote $\mathbf{L=}\frac{\partial }{\partial y}-\mathbf{M}$ and $%
\mathbf{D}=\frac{\partial }{\partial t}-\mathbf{A}$. \ 

\begin{lemma}
\textit{Let }$\mathbf{\psi }$\textit{\ be a solution of the system (1.1),
whose the coefficients }$\mathbf{q}_{12}$\textit{\ and }$\mathbf{q}_{21}$%
\textit{\ satisfy system (2.3). Then the function }$\mathbf{\varphi }=%
\mathbf{D\psi }$\textit{\ also satisfy the system (1.1).}

\begin{proof}
From (2.5) we obtain: 
\begin{equation*}
(\mathbf{LD}-\mathbf{DL)\psi }=\mathbf{L(D\psi )-D(\mathbf{L}\psi )=0.}
\end{equation*}%
Since $\mathbf{\mathbf{L}\psi }=0$, then 
\begin{equation*}
\mathbf{L(D}\psi )=0.
\end{equation*}%
It means that $\mathbf{D}\psi $\textbf{\ }is solution of system (1.1).
\end{proof}
\end{lemma}

\section{\textbf{Manakov-type systems on the plane}}

\bigskip Let us consider the system (1.1) on the plane $-\infty <x,y<+\infty 
$ with the matrix function $\mathbf{q}_{12}$ and $\mathbf{q}_{21}$ has
measurable complex-valued rapidly decreasing (Schwartz) entries. Notice that
if the potential is independent on $y$, then by taking $\mathbf{\psi }(x,y)=%
\mathbf{\psi }(x)exp(i\lambda y)$, we can convert equation (1.1) into the
Manakov system given by

\begin{equation*}
-\mathbf{\sigma }\frac{d}{dx}\mathbf{\psi }(x)+\mathbf{Q}(x)\mathbf{\psi }%
(x)=i\lambda \mathbf{\psi }(x)
\end{equation*}%
which is considered in [10].

Throughout this chapter the following notations will be used :

$\blacktriangleright \ $ We partition $\left( n+1\right) \times \left(
n+1\right) $ matrix\textbf{\ }$\mathbf{A}$ as follows:%
\begin{equation*}
\mathbf{A}=\left( 
\begin{array}{cc}
\mathbf{A}_{11} & \mathbf{A}_{12} \\ 
\mathbf{A}_{21} & A_{22}%
\end{array}%
\right)
\end{equation*}%
where $\mathbf{A}_{11}$ is $n\times n$ matrix $\mathbf{A}_{12}$ is and $%
n\times 1$ column vector, $\mathbf{A}_{21}$ is $1\times n$ row vector and $%
A_{22}$ is scalar.

$\blacktriangleright $ $\mathbf{\digamma }_{x}$ denotes the $\left(
n+1\right) \times \left( n+1\right) $ diagonal matrix shift operator, such
that for a $n+1$ dimensional vector function $\mathbf{h}\left( t\right) $ 
\begin{equation*}
\mathbf{\ \digamma }_{x}\mathbf{h}\left( y\right) =\left( 
\begin{array}{c}
\mathbf{h}_{1}\left( y+x\right) \\ 
h_{2}\left( y-x\right)%
\end{array}%
\right)
\end{equation*}%
where $\mathbf{h}_{1}\left( y\right) $ is a vector function that consists of
the first $n$ component of vector $h\left( y\right) ,$ $h_{2}\left( y\right) 
$ is a scalar function.

$\blacktriangleright $ We denote

\begin{equation*}
\mathbf{A}_{\pm }\left( x\right) h\left( x,y\right) =\mp \int_{y}^{\mp
\infty }\mathbf{A}_{\pm }\left( x,y,\tau \right) h\left( x,\tau \right) d\tau
\end{equation*}%
by the upper and lower Volterra integral operators.

\bigskip

\bigskip \textit{For any }$\mathbf{a}_{\pm }\left( y\right) \in \mathbf{L}%
_{2}\left( 
\mathbb{R}
,%
\mathbb{C}
^{n+1}\right) $\textit{\ there exist unique solutions in }$\mathbf{L}%
_{2}\left( 
\mathbb{R}
^{2},%
\mathbb{C}
^{n+1}\right) $\textit{\ of the systems (1.1) with the conditions }\textbf{\ 
}$\mathbf{\psi }\left( x,y\right) =\mathbf{\digamma }_{x}\mathbf{a}_{\pm
}\left( y\right) +o(1),\ \ y\longrightarrow \pm \infty $\textit{\ and these
solutions admit the representations }%
\begin{equation}
\mathbf{\psi }\left( x,y\right) =\left( \mathbf{I+A}_{\pm }\left( x\right)
\right) \mathbf{\digamma }_{x}\mathbf{a}_{\mp }\left( y\right) ,  \tag{3.1}
\end{equation}%
\textit{where }$\mathbf{I}$ \textit{is identity operator and the }kernels $%
\mathbf{A}^{\pm }\left( x,y,\tau \right) =$\ $\left( 
\begin{array}{cc}
\mathbf{A}_{11}^{\pm }\left( x,y,\tau \right) & \mathbf{A}_{12}^{\pm }\left(
x,y,\tau \right) \\ 
\mathbf{A}_{21}^{\pm }\left( x,y,\tau \right) & A_{22}^{\pm }\left( x,y,\tau
\right)%
\end{array}%
\right) $ of the integral operators $\mathbf{A}_{\pm }\left( x\right) $%
\textit{\ are uniquely determined by the coefficients of the system (1.1),
and for the fixed }$x$\textit{, these kernels are the Hilbert-Schmidt
kernels. In addition, these kernels are connected with the potential by the
following equalities}

\begin{equation}
\mathbf{A}_{12}^{\pm }\left( x,y,y\right) =\pm \frac{1}{2}\mathbf{q}%
_{12}\left( x,y\right) ,\text{ }\mathbf{A}_{21}^{\pm }\left( x,y,y\right)
=\mp \frac{1}{2}\mathbf{q}_{21}\left( x,y\right) .  \tag{3.2}
\end{equation}

An operator $\mathbf{S}$ transforming the given incident waves $\mathbf{a}%
_{-}\left( y\right) \in \mathbf{L}_{2}\left( 
\mathbb{R}
,%
\mathbb{C}
^{n+1}\right) $ into the scattered waves $\mathbf{a}_{+}\left( y\right) \in 
\mathbf{L}_{2}\left( 
\mathbb{R}
,%
\mathbb{C}
^{n+1}\right) $ is called the scattering operator for the system (1) on the
plane: 
\begin{equation*}
\mathbf{a}_{+}\left( y\right) =\mathbf{Sa}_{-}\left( y\right)
\end{equation*}%
where $\mathbf{a}_{+}\left( y\right) =\mathbf{a}_{-}\left( y\right)
+\int_{-\infty }^{+\infty }\mathbf{\digamma }_{y-x-s}\left( \mathbf{Q\psi }%
\right) \left( x,s\right) ds$. Operator\textbf{\ }$\mathbf{S}$ is $\left(
n+1\right) \times \left( n+1\right) $ matrix linear operator on the space $%
\mathbf{L}_{2}\left( 
\mathbb{R}
,%
\mathbb{C}
^{n+1}\right) $.

From the representations (3.1) it follows that the next factorization
results for $\mathbf{S}$. For every $x,$ the operator $\mathbf{\digamma }_{x}%
\mathbf{S\digamma }_{-x}$ admits factorizations 
\begin{equation}
\mathbf{\digamma }_{x}\mathbf{S\digamma }_{-x}=\left( \mathbf{I+A}_{-}\left(
x\right) \right) ^{-1}\left( \mathbf{I+A}_{+}\left( x\right) \right) , 
\tag{3.3}
\end{equation}

\bigskip We can analogously introduce the next representations corresponding
to asymptotics $\mathbf{\psi }\left( x,y\right) =\mathbf{\digamma }_{x}%
\mathbf{b}_{-}\left( y\right) +o(1),\ \ x\longrightarrow -\infty \ $and $%
\mathbf{\psi }\left( x,y\right) =\mathbf{\digamma }_{x}\mathbf{b}_{+}\left(
y\right) +o(1),\ \ x\longrightarrow +\infty $.

\textit{For any }$\mathbf{b}_{\pm }\left( y\right) \in \mathbf{L}_{2}\left( 
\mathbb{R}
,%
\mathbb{C}
^{n}\right) $\textit{\ there exist unique solutions in }$\mathbf{L}%
_{2}\left( 
\mathbb{R}
^{2},%
\mathbb{C}
^{n}\right) $\textit{\ of the systems () and these solutions admit the
representations }%
\begin{equation}
\mathbf{\psi }\left( x,y\right) =\left( \mathbf{I+B}_{\pm }\left( y\right)
\right) \mathbf{\digamma }_{x}\mathbf{b}_{\mp }\left( y\right) ,  \tag{3.4}
\end{equation}%
\textit{where the kernels }$\mathbf{B}^{\pm }\left( x,y,\tau \right) =$%
\textit{\ }$\left( 
\begin{array}{cc}
\mathbf{B}_{11}^{\pm }\left( x,y,\tau \right) & \mathbf{B}_{12}^{\pm }\left(
x,y,\tau \right) \\ 
\mathbf{B}_{21}^{\pm }\left( x,y,\tau \right) & B_{22}^{\pm }\left( x,y,\tau
\right)%
\end{array}%
\mathit{\ }\right) $ \textit{of the integral operators }$\mathbf{B}_{\pm
}\left( x\right) $\textit{\ are uniquely determined by the coefficients of
the system (1.1) and these kernels are the Hilbert-Schmidt kernels for fixed 
}$x$\textit{. In addition, these kernels are connected with the potential by
the following equalities}

\begin{equation}
\mathbf{B}_{12}^{\pm }\left( x,y,x\right) =\pm \frac{1}{2}\mathbf{q}%
_{12}\left( x,y\right) ,\text{ }\mathbf{B}_{21}^{\pm }\left( x,y,x\right)
=\mp \frac{1}{2}\mathbf{q}_{21}\left( x,y\right) .  \tag{3.5}
\end{equation}

\bigskip From the representation (3.4) follows the next factorization
results for $\mathbf{S}$. For every $x,$ the operator $\mathbf{\digamma }_{x}%
\mathbf{S\digamma }_{-x}$ admits factorizations 
\begin{equation}
\mathbf{\digamma }_{x}\mathbf{S\digamma }_{-x}=\left( \mathbf{I+K}_{+}\left(
x\right) \right) ^{-1}\left( \mathbf{I+K}_{-}\left( x\right) \right) 
\tag{3.6}
\end{equation}%
where matrix elements of the kernel of matrix integral operator $\mathbf{K}%
_{\pm }\left( x\right) $ are determined by $\mathbf{B}_{\pm }\left( y\right) 
$ as follows%
\begin{eqnarray}
\mathbf{K}_{i1}^{\pm }\left( x,y,\tau \right) &=&\mathbf{B}_{i1}^{\pm
}\left( x,y,x-y+\tau \right) ,  \TCItag{3.7} \\
\mathbf{K}_{i2}^{\pm }\left( x,y,\tau \right) &=&\mathbf{B}_{i2}^{\pm
}\left( x,y,x+y-\tau \right) ,i=1,2.  \notag
\end{eqnarray}

It is possible the unique restoration of the potential by scattering
operator.

Let $\mathbf{S}$ be the scattering operator for the system (1.1) on the
plane with the potential\textbf{\ }$\mathbf{Q}\left( x,y\right) $ belonging
to the Schwartz class. Then the potential $\mathbf{Q}\left( x,y\right) $ is
uniquely determined by the known scattering operator $\mathbf{S}$. The ISP
is solved with the following steps:

1) Construct the operator $\mathbf{\digamma }_{x}\mathbf{S\digamma }_{-x}$;

2) Find the factorization factors $\mathbf{A}_{-}\left( x\right) $ and $%
\mathbf{A}_{+}\left( x\right) $ from the (3.3), since $\mathbf{\digamma }_{x}%
\mathbf{S\digamma }_{-x}$ admits left factorization;

3) Find the matrix coefficients of the system (1.1) with respect to the
kernels $\mathbf{A}^{\pm }\left( x,y,\tau \right) $ of the operators $%
\mathbf{A}_{\pm }\left( x\right) $, by formula (3.2).

Let us denote the kernels of the matrix integral operators $\mathbf{S-I}$
and $\mathbf{S}^{-1}\mathbf{-I}$ as $\mathbf{F}\left( y,\tau \right) $ and $%
\mathbf{G}\left( y,\tau \right) $, respectively. Let$\ \mathbf{F}\left(
y,\tau \right) =\left( 
\begin{array}{cc}
\mathbf{F}_{11}\left( y,\tau \right) & \mathbf{F}_{12}\left( y,\tau \right)
\\ 
\mathbf{F}_{21}\left( y,\tau \right) & F_{22}\left( y,\tau \right)%
\end{array}%
\right) $ and $\mathbf{G}\left( t,\tau \right) =\left( 
\begin{array}{cc}
\mathbf{G}_{11}\left( y,\tau \right) & \mathbf{G}_{12}\left( y,\tau \right)
\\ 
\mathbf{G}_{21}\left( y,\tau \right) & G_{22}\left( y,\tau \right)%
\end{array}%
\right) $. We will call the collection of functions $\left\{ \mathbf{F}%
_{12}\left( y,\tau \right) ,\mathbf{G}_{21}\left( y,\tau \right) \right\} $\
as the scattering data for the system (1.1)$.$

Let us denote the kernels of the matrices $\mathbf{\digamma }_{x}\mathbf{%
S\digamma }_{-x}\mathbf{-I}$ and $\mathbf{\digamma }_{x}\mathbf{S}^{-1}%
\mathbf{\digamma }_{-x}-\mathbf{I}$ as $\mathbf{F}\left( x,y,\tau \right) $
and $\mathbf{G}\left( x,y,\tau \right) $, respectively. It is clear that $%
\mathbf{F}\left( 0,y,\tau \right) =$ $\mathbf{F}\left( y,\tau \right) $ and $%
\mathbf{G}\left( 0,y,\tau \right) =\mathbf{G}\left( y,\tau \right) $.

Therefore, we obtain the following Gelfand- Levitan-Marchenko type matrix
integral equations from (3.6).

\begin{eqnarray}
\mathbf{K}_{12}^{-}\left( x,y,\tau \right) -\int_{-\infty }^{y}\left[
\int_{y}^{+\infty }\mathbf{K}_{12}^{-}\left( x,y,z\right) \mathbf{G}%
_{21}\left( z-x,s+x\right) dz\right] \mathbf{F}_{12}\left( s+x,\tau
-x\right) ds &=&  \notag \\
\mathbf{F}_{12}\left( y+x,\tau -x\right) ,\text{ }\tau &\geq &y,  \notag
\end{eqnarray}%
\begin{equation}
\tag{3.8}
\end{equation}

\begin{eqnarray}
\mathbf{K}_{21}^{+}\left( x,y,\tau \right) -\int_{y}^{+\infty }\left[
\int_{-\infty }^{y}\mathbf{K}_{21}^{+}\left( x,y,z\right) \mathbf{F}%
_{12}\left( z+x,s-x\right) dz\right] \mathbf{G}_{21}\left( s-x,\tau
+x\right) ds &=&  \notag \\
\mathbf{G}_{21}\left( y-x,\tau +x\right) ,\text{ }\tau &\leq &y,  \notag
\end{eqnarray}

By the right factorization (3.6) of $\mathbf{\digamma }_{x}\mathbf{S\digamma 
}_{-x}$ there exist unique solutions of these equations.

Considering the relationships (3.5) between the potential $\mathbf{Q}\left(
x,t\right) $ and the operators $\mathbf{B}_{\pm }\left( y\right) $, and also
(3.7) between the operators $\mathbf{B}_{\pm }\left( y\right) $ and $\mathbf{%
K}_{\pm }\left( x\right) $ we obtain the following results for the ISP in
the plane:

\textit{Let }$\mathbf{S=I+F}$\textit{\ be the scattering operator for the
system (1.1) on the whole plane. Then there exists }$\mathbf{S}^{-1}\mathbf{%
=I+G}$\textit{, where }$\mathbf{F}$\textbf{\ }\textit{and }$\mathbf{G}$%
\textit{\ are the Hilbert-Schmidt matrix integral operators. Let us
partition }$\mathbf{F=}\left( \mathbf{F}_{ij}\right) _{i,j=1}^{2},$\textit{\ 
}$\mathbf{G}=\left( \mathbf{G}_{ij}\right) _{i,j=1}^{2}$\textit{\ and let
the kernels of the operators }$\mathbf{F}_{12}$ and $\mathbf{G}_{21}$\textit{%
\ be given. Then there exists a unique solution of the system of integral
equations (3.8) and the solution of this system determines the potential by
formulae}%
\begin{equation}
\mathbf{q}_{12}\left( x,y\right) =-2\mathbf{K}_{12}^{-}\left( x,y,y\right)
,\ \mathbf{q}_{21}\left( x,y\right) =-2\mathbf{K}_{21}^{+}\left(
x,y,y\right) .  \tag{3.9}
\end{equation}

\bigskip Thus, for the system (1.1) with the coefficient $\mathbf{Q}(x,y)$
there is scattering operator $\mathbf{S}$ with the scattering data $\mathbf{F%
}_{12}$ and $\mathbf{G}_{21}$ which are the Hilbert-Schmidt integral
operators with the kernels $\mathbf{F}_{12}\left( x,y\right) $ and $\mathbf{G%
}_{21}\left( x,y\right) $ that decrease quite fast with respect to variables
at infinity. This defines the mapping of the scattering data%
\begin{equation*}
\left\{ \mathbf{q}_{12}(x,y),\mathbf{q}_{21}^{T}(x,y)\right\} \overset{\Pi }{%
\rightarrow }\left\{ \mathbf{F}_{12}\left( x,y\right) ,\mathbf{G}%
_{21}^{T}\left( x,y\right) \right\} .
\end{equation*}%
This operator mapping coefficients of the system (1.1) into the scattering
data is continuous in $L_{2}$ and its inverse $\Pi ^{-1}$ exists and is
continuous and its action can be constructively described by means of the
uniquely solvable of systems (3.8).

\section{\textbf{Inverse scattering method}}

To integrate the Cauchy problem for the system (2.1) with the initial
condition 
\begin{equation}
\left. q_{k}(x,y,t)\right\vert _{t=0}=q_{k}^{0}(x,y),k=1,...,2n  \tag{4.1}
\end{equation}%
by the inverse scattering method. We use the ISP for the system (1.1) with
the potential $\mathbf{Q}^{0}\mathbf{(}x,y\mathbf{)}=\left( 
\begin{array}{cc}
\mathbf{0}_{n} & \mathbf{q}_{12}^{0}\mathbf{(}x,y\mathbf{)} \\ 
\mathbf{q}_{21}^{0}\mathbf{(}x,y\mathbf{)} & 0%
\end{array}%
\right) $, where $\mathbf{q}_{12}^{0}=\func{col}\left\{
q_{1}^{0},...,q_{n}^{0}\right\} $, $\mathbf{q}_{21}^{0}=\func{col}\left\{
q_{n+1}^{0},...,q_{2n}^{0}\right\} ^{T}$ in the whole plane, that is given
in Chapter 3. Let $\mathbf{F}_{12}^{0}\left( x,y\right) $ and $\mathbf{G}%
_{21}^{0}\left( x,y\right) $ are the scattering data for the system (1.1)
with the coefficient $\mathbf{Q}^{0}\mathbf{(}x,y\mathbf{)}$ which decrease
quite fast with respect to variables at infinity. This defines the mapping
of the scattering data%
\begin{equation*}
\mathbf{q}^{0}=\left[ 
\begin{array}{c}
\mathbf{q}_{12}^{0} \\ 
\left( \mathbf{q}_{21}^{0}\right) ^{T}%
\end{array}%
\right] \overset{\Pi }{\rightarrow }\left[ 
\begin{array}{c}
\mathbf{F}_{12}^{0} \\ 
\left( \mathbf{G}_{21}^{0}\right) ^{T}%
\end{array}%
\right] .
\end{equation*}

Let us investigate the evolution of this scattering data, when the
coefficients of the operator $\mathbf{L}$ satisfy the equations (2.3).

The pair $\left\{ \mathbf{F}_{12},\mathbf{G}_{21}\right\} $ is denoted the
scattering data correspond to operator $\mathbf{L}$ with the coefficients $%
\mathbf{q}_{12}(x,y;t)$\ and $\mathbf{q}_{21}(x,y;t)$\ which are satisfy the
system of equation (2.3).

\begin{theorem}
\textit{Let the coefficients }$\mathbf{q}_{12}$\textit{\ and }$\mathbf{q}%
_{21}$\textit{\ of the system (1.1) depend on }$t$\textit{\ as a parameter
and satisfy the system of equation (2.3). Besides that }%
\begin{equation*}
\mathbf{P}(x,+\infty )=\mathbf{0},\text{ \ }p(x,-\infty )=0^{-}.
\end{equation*}%
\textit{Then the kernels}\textbf{\ }$\mathbf{F}_{12}(y,\tau ;t),\mathbf{G}%
_{21}(y,\tau ;t)$\textit{\ of the integral operators }$\mathbf{F}_{12},%
\mathbf{G}_{21}$\textit{\ corresponding to the scattering operator }$\mathbf{%
S}$\textit{\ for the system (1.1) on the plane satisfy the system of
equations (4.5)\ and (4.6).}

\begin{proof}
By virtue of definition of the scattering operator $\mathbf{S}$ we get%
\begin{equation}
\mathbf{\varphi }_{+}=\mathbf{S\varphi }_{-},  \tag{4.2}
\end{equation}%
where $\mathbf{\varphi }_{\pm }=\mathbf{P}_{\pm }\mathbf{a}_{\pm }$, $%
\mathbf{P}_{\pm }=\frac{\partial }{\partial t}-\mathbf{\delta \sigma }\frac{%
\partial }{\partial y}-i\gamma \mathbf{I}_{n}\frac{\partial ^{2}}{\partial
y^{2}},$ $\mathbf{P}(x,$ $\pm \infty )=\mathbf{0}$, $p(x,$ $\pm \infty )=0$.
Since $\mathbf{a}_{+}=\mathbf{Sa}_{-}$, from (4.2) we obtain: 
\begin{equation}
\mathbf{P}_{+}\mathbf{S=SP}_{-}.  \tag{4.3}
\end{equation}%
Analogously, 
\begin{equation}
\mathbf{P}_{-}\mathbf{S}^{-1}\mathbf{=S}^{-1}\mathbf{P}_{+}\mathbf{.} 
\tag{4.4}
\end{equation}%
Since $\mathbf{S=I+F}$ and $\mathbf{S}^{-1}\mathbf{=I+G}$, where $\mathbf{F}%
f(y)=\dint\limits_{-\infty }^{+\infty }\mathbf{F}(y,\tau ;t)f(\tau )d\tau ,$ 
$\mathbf{F}(y,\tau ;t)=\left( \mathbf{F}_{ij}(y,\tau ;t)\right) _{i,j=1}^{2}$
and $\mathbf{G}f(y)=\dint\limits_{-\infty }^{+\infty }\mathbf{G}(y,\tau
;t)f(\tau )d\tau ,$ $\mathbf{G}(y,\tau ;t)=\left( \mathbf{G}_{ij}(y,\tau
;t)\right) _{i,j=1}^{2},$ from the matrix operator equation (4.3) it follows
that the kernels of the integral operator $\mathbf{F}_{12}$ satisfy the
equation 
\begin{equation}
\frac{\partial }{\partial t}\mathbf{F}_{12}-2\left( \mathbf{B}\frac{\partial 
}{\partial y}\mathbf{F}_{12}-b\frac{\partial }{\partial \tau }\mathbf{F}%
_{12}\right) -i\gamma \left( \frac{\partial ^{2}}{\partial y^{2}}\mathbf{F}%
_{12}-\frac{\partial ^{2}}{\partial \tau ^{2}}\mathbf{F}_{12}\right) =%
\mathbf{0},  \tag{4.5}
\end{equation}%
The similarly equation for the kernels of the integral operator $\mathbf{G}%
_{21}$ follows from the matrix operator equation (4.4): 
\begin{equation}
\frac{\partial }{\partial t}\mathbf{G}_{21}+2\left( b\frac{\partial }{%
\partial y}\mathbf{G}_{21}-\frac{\partial }{\partial \tau }\mathbf{G}_{21}%
\mathbf{B}\right) +i\gamma \left( \frac{\partial ^{2}}{\partial y^{2}}%
\mathbf{G}_{21}-\frac{\partial ^{2}}{\partial \tau ^{2}}\mathbf{G}%
_{21}\right) =0.  \tag{4.6}
\end{equation}
\end{proof}
\end{theorem}

Now, let us give a procedure for the solution of the system (2.3) by inverse
scattering method.

\begin{theorem}
Let \textit{functions }$\mathbf{F}_{12}(y,\tau ;t)$ and $\mathbf{G}%
_{21}(y,\tau ;t)$ satisfy the equations (4.5) and (4.6) and these functions
together with their derivatives with respect to t and their first and second
derivatives with respect to $y$ and $\tau $ belong to $\mathbf{L}_{2}(%
\mathbb{R}
^{2})$. Then the equations (4.7) are uniquely solvable and \textit{the
functions }%
\begin{equation*}
\mathbf{q}_{12}(x,y;t)=-2\mathbf{K}^{-}(x,y,y;t),\text{ \ }\mathbf{q}%
_{21}(x,y;t)=-2\mathbf{K}^{+}(x,y,y;t),
\end{equation*}%
\textit{is the solution of the non linear equation (2.3). }

\begin{proof}
If the coefficients of the system (1.1) depend on $t$ as a parameter and
satisfy the system of equation (2.3), then the kernels $\mathbf{F}%
_{12}(y,\tau ;t),\mathbf{G}_{21}(y,\tau ;t)$ of the integral operators $%
\mathbf{F}_{12}\mathbf{,G}_{21}$ satisfy the system of equations (4.5),
(4.6). In addition, the coefficients $\mathbf{q}_{12}$, $\mathbf{q}_{21}$ of
the system (1.1) is uniquely determined by (3.9) and the analogues of the
equations (3.8) 
\begin{eqnarray}
&&%
\begin{array}{c}
\mathbf{K}^{-}\left( x,y,\tau ;t\right) -\int_{-\infty }^{y}\mathbf{K}%
^{-}\left( x,y,z;t\right) \left( \int_{y}^{+\infty }\mathbf{G}_{21}\left(
z-x,s+x;t\right) \mathbf{F}_{12}\left( s+x,\tau -x;t\right) ds\right) dz \\ 
=\mathbf{F}_{12}\left( y+x,\tau -x;t\right) ,\tau \geq y,%
\end{array}
\notag \\
&&  \TCItag{4.7} \\
&&%
\begin{array}{c}
\mathbf{K}^{+}\left( x,y,\tau ;t\right) -\int_{y}^{+\infty }\mathbf{K}%
^{+}\left( x,y,z;t\right) \left( \int_{-\infty }^{y}\mathbf{F}_{12}\left(
z+x,s-x;t\right) \mathbf{G}_{21}\left( s-x,\tau +x;t\right) ds\right) dz \\ 
=\mathbf{G}_{21}\left( y-x,\tau +x;t\right) ,\tau \leq y,%
\end{array}
\notag
\end{eqnarray}%
which are constructed by kernels $\mathbf{F}_{12}(y,\tau ;t),\mathbf{G}%
_{21}(y,\tau ;t)$ are uniquely solved by $\mathbf{K}^{-}\left( x,y,\tau
\right) ,\mathbf{K}^{+}\left( x,y,\tau \right) $
\end{proof}
\end{theorem}

The statement of this theorem is equivalent to the assertion that the
function 
\begin{equation}
\mathbf{q}=\left[ 
\begin{array}{c}
\mathbf{q}_{12} \\ 
\left( \mathbf{q}_{21}\right) ^{T}%
\end{array}%
\right] =\mathbf{\Pi }^{-1}e^{-i\mathbf{A}t}\mathbf{\Pi q}^{0}  \tag{4.8}
\end{equation}%
is the solution of equation (2.3) with the initial condition (4.1) if it is
determined, where 
\begin{equation*}
\mathbf{A}=\left[ 
\begin{array}{cc}
\mathbf{A}_{1} & \mathbf{0}_{n} \\ 
\mathbf{0}_{n} & \mathbf{A}_{2}%
\end{array}%
\right] ,\left[ 
\begin{array}{c}
\mathbf{A}_{1}=2i\left( \mathbf{B}\frac{\partial }{\partial y}-b\mathbf{I}%
_{n}\frac{\partial }{\partial \tau }\right) -\gamma \mathbf{I}_{n}\left( 
\frac{\partial ^{2}}{\partial y^{2}}-\frac{\partial ^{2}}{\partial \tau ^{2}}%
\right) \\ 
\mathbf{A}_{2}=2i\left( \mathbf{B}\frac{\partial }{\partial \tau }\mathbf{-}b%
\mathbf{I}_{n}\frac{\partial }{\partial y}\right) +\gamma \mathbf{I}%
_{n}\left( \frac{\partial ^{2}}{\partial y^{2}}-\frac{\partial ^{2}}{%
\partial \tau ^{2}}\right) .%
\end{array}%
\right]
\end{equation*}%
It is well known that $\mathbf{\Pi q}^{0}=\left[ 
\begin{array}{c}
\mathbf{F}_{12}^{0} \\ 
\left( \mathbf{G}_{21}^{0}\right) ^{T}%
\end{array}%
\right] $ is direct problem of determining scattering data, $e^{-i\mathbf{A}%
t}\mathbf{\Pi q}^{0}$ is the evolution of scattering data and $\mathbf{\Pi }%
^{-1}e^{-i\mathbf{A}t}\mathbf{\Pi q}^{0}$ is the inverse scattering problem
of finding $\mathbf{q}$.

The function $\mathbf{q}(x,y;t)$ these functions together with their
derivatives with respect to $t$ and their first and second derivatives with
respect to $x$ and $y$ belong to $\mathbf{L}_{2}(%
\mathbb{R}
^{2})$ is called the solution of Cauchy problem (1.1), (4.1) if at $t=0$ it
coincides with the initial data $\mathbf{q(}x,y,0\mathbf{)=q}^{0}(x,y)$.

\begin{theorem}
The solution of Cauchy problem (1.1), (1.4) is unique. The solution of this
Cauchy problem exists on an arbitrary interval of time for small initial
data.

\begin{proof}
The uniqueness of the solution follows from the possibility of representing
it in the form (4.8) that the assumption of the existence of solution
requires the representation (4.8) which is expressed by initial data. If the
initial data $\mathbf{q}^{0}$ are sufficiently small in (4.8) then this
formula has a sense by virtue of continuity of $\mathbf{\Pi }$ and unitarity
of $e^{-i\mathbf{A}t}$ that $\left\Vert e^{-i\mathbf{A}t}\mathbf{\Pi q}%
^{0}\right\Vert $ is less than 1.
\end{proof}
\end{theorem}

\section{\protect\bigskip \textbf{Exact soliton-like solutions of the
dispersive }$4$\textbf{-wave interaction problem}}

Let $n=2$ in (1.1) and 
\begin{equation}
\mathbf{Q}=\left( 
\begin{array}{cc}
0 & \mathbf{q}_{12} \\ 
\mathbf{q}_{21} & 0%
\end{array}%
\right) ,\text{ }\mathbf{q}_{12}=\left( 
\begin{array}{c}
q_{1} \\ 
q_{2}%
\end{array}%
\right) ,\text{ \ }\mathbf{q}_{21}=\left( 
\begin{array}{cc}
q_{3} & q_{4}%
\end{array}%
\right) .  \tag{5.1}
\end{equation}
It is easy to see that the scattering data corresponding to the potential
(5.1) is in the form of 
\begin{equation*}
\mathbf{F}_{12}\left( y,\tau ;t\right) =\left( 
\begin{array}{c}
f_{11}\left( y,\tau ;t\right) \\ 
f_{21}\left( y,\tau ;t\right)%
\end{array}%
\right) ,\mathbf{G}_{21}\left( y,\tau ;t\right) =\left( 
\begin{array}{cc}
g_{11}\left( y,\tau ;t\right) & g_{12}\left( y,\tau ;t\right)%
\end{array}%
\right) .
\end{equation*}

The nonlinear system of equations (2.1) becomes to the form 
\begin{eqnarray}
\partial _{t}q_{1}+\alpha _{1}\partial _{x}q_{1}+\beta _{1}\partial
_{y}q_{1}-i\gamma \partial _{xy}^{2}q_{1} &=&\left( p_{11}-p\right)
q_{1}+p_{12}q_{2},  \notag \\
\partial _{t}q_{2}+\alpha _{2}\partial _{x}q_{2}+\beta _{2}\partial
_{y}q_{2}-i\gamma \partial _{xy}^{2}q_{2} &=&p_{21}q_{1}+\left(
p_{22}-p\right) q_{2},  \notag \\
&&  \TCItag{5.2} \\
\partial _{t}q_{3}+\alpha _{1}\partial _{x}q_{3}+\beta _{1}\partial
_{y}q_{3}+i\gamma \partial _{xy}^{2}q_{3} &=&\left( p-p_{11}\right)
q_{3}-p_{21}q_{4},  \notag \\
\partial _{t}q_{4}+\alpha _{2}\partial _{x}q_{4}+\beta _{2}\partial
_{y}q_{4}+i\gamma \partial _{xy}^{2}q_{4} &=&\left( p-p_{22}\right)
q_{4}-p_{12}q_{3},  \notag
\end{eqnarray}%
where%
\begin{eqnarray}
\frac{\partial }{\partial \xi }p &=&-i\gamma \frac{\partial }{\partial x}%
(q_{1}q_{3})-i\gamma \frac{\partial }{\partial x}(q_{2}q_{4}),  \notag \\
\text{ }\frac{\partial }{\partial \eta }p_{11} &=&-i\gamma \frac{\partial }{%
\partial x}(q_{1}q_{3}),\text{ }\frac{\partial }{\partial \eta }%
p_{22}=-i\gamma \frac{\partial }{\partial x}(q_{2}q_{4}),  \TCItag{5.3} \\
\frac{\partial }{\partial \eta }p_{km} &=&\left( \beta _{m}-\beta
_{k}\right) q_{k}q_{2+m}-i\gamma \frac{\partial }{\partial x}(q_{k}q_{2+m}),%
\text{ }m,k=1,2;\text{ }m\neq k.  \notag
\end{eqnarray}

In the case $P^{+}(x)=p^{-}(x)=0$, these derivatives comes to form 
\begin{eqnarray}
p &=&i\frac{\gamma }{2}(q_{1}q_{3})+i\frac{\gamma }{2}(q_{2}q_{4})-i\frac{%
\gamma }{2}\dint\limits_{-\infty }^{\xi }\left[ \frac{\partial }{\partial
\eta }(q_{1}q_{3})+\frac{\partial }{\partial \eta }(q_{2}q_{4})\right] ds, 
\notag \\
\text{ }p_{11} &=&i\frac{\gamma }{2}(q_{1}q_{3})-i\frac{\gamma }{2}%
\dint\limits_{\eta }^{+\infty }\frac{\partial }{\partial \zeta }%
(q_{1}q_{3})d\tau ,\text{ }p_{22}=i\frac{\gamma }{2}(q_{2}q_{4})-i\frac{%
\gamma }{2}\dint\limits_{\eta }^{+\infty }\frac{\partial }{\partial \zeta }%
(q_{2}q_{4})d\tau ,  \TCItag{5.4} \\
p_{km} &=&\frac{\beta _{m}-\beta _{k}}{2}q_{k}q_{2+m}+i\frac{\gamma }{2}%
(q_{k}q_{2+m})-i\frac{\gamma }{2}\dint\limits_{\eta }^{+\infty }\frac{%
\partial }{\partial \zeta }(q_{k}q_{2+m})d\tau ,\text{ }m,k=1,2;\text{ }%
m\neq k.  \notag
\end{eqnarray}%
and after the elimination of $p$ and $p_{km}$, the system (5.2) represents a
system of integro-differential equations.

The evolution of the scattering data are in the following form according to
(4.5) and (4.6):%
\begin{eqnarray}
\partial _{t}f_{11}-2b_{1}\partial _{y}f_{11}+2b\partial _{\tau
}f_{11}-i\gamma \left( \partial _{y}^{2}f_{11}-\partial _{\tau
}^{2}f_{11}\right) &=&0,  \notag \\
\partial _{t}f_{21}+2b_{2}\partial _{y}f_{21}-2b\partial _{\tau
}f_{21}-i\gamma \left( \partial _{y}^{2}f_{21}-\partial _{\tau
}^{2}f_{21}\right) &=&0,  \notag \\
&&  \TCItag{5.5} \\
\partial _{t}g_{11}+2b\partial _{y}g_{11}-2b_{1}\partial _{\tau
}g_{11}+i\gamma \left( \partial _{y}^{2}f_{11}-\partial _{\tau
}^{2}f_{11}\right) &=&0,  \notag \\
\partial _{t}g_{12}+2b\partial _{y}g_{12}-2b_{2}\partial _{\tau
}g_{12}+i\gamma \left( \partial _{y}^{2}f_{21}-\partial _{\tau
}^{2}f_{21}\right) &=&0.  \notag
\end{eqnarray}

We deduce explicit solutions of the system (5.2) by using the formulas for
the exactly solvable case of the inverse-scattering problem for the system
(1.1). We get an elementary example for $F_{12}\left( y,\tau \right) =\left( 
\begin{array}{c}
f_{1}\left( y;t\right) f_{2}\left( \tau ;t\right) \\ 
f_{3}\left( y;t\right) f_{4}\left( \tau ;t\right)%
\end{array}%
\right) $, $G_{21}\left( y,\tau \right) =\left( 
\begin{array}{cc}
g_{1}\left( y;t\right) g_{2}\left( \tau ;t\right) & g_{3}\left( y;t\right)
g_{4}\left( \tau ;t\right)%
\end{array}%
\right) $, where the functions $f_{k}$ and $g_{k}$ satisfy the equations%
\begin{eqnarray}
\partial _{t}f_{1}-2b_{1}\partial _{y}f_{1}-i\gamma \partial _{y}^{2}f_{1}
&=&0,\text{ }\partial _{t}f_{2}+2b\partial _{\tau }f_{2}+i\gamma \partial
_{\tau }^{2}f_{2}=0,\text{ }  \notag \\
\text{ }\partial _{t}f_{3}-2b_{2}\partial _{y}f_{3}-i\gamma \partial
_{y}^{2}f_{3} &=&0,\text{ }\partial _{t}f_{4}+2b\partial _{\tau
}f_{4}+i\gamma \partial _{\tau }^{2}f_{4}=0,  \notag \\
&&  \TCItag{5.6} \\
\partial _{t}g_{1}+2b_{1}\partial _{y}g_{1}+i\gamma \partial _{y}^{2}g_{1}
&=&0,\text{ }\partial _{t}g_{2}+2b_{1}\partial _{\tau }g_{2}-i\gamma
\partial _{\tau }^{2}g_{2}=0,\text{ }  \notag \\
\text{ }\partial _{t}g_{3}+2b\partial _{y}g_{3}+i\gamma \partial
_{y}^{2}g_{3} &=&0,\text{ }\partial _{t}g_{4}-2b_{2}\partial _{\tau
}g_{4}-i\gamma \partial _{\tau }^{2}g_{4}=0.  \notag
\end{eqnarray}

\bigskip Let $\mathbf{K}_{12}^{-}=\left[ 
\begin{array}{c}
K_{1}^{-} \\ 
K_{2}^{-}%
\end{array}%
\right] ,\mathbf{K}_{21}^{+}=\left[ 
\begin{array}{cc}
K_{1}^{+} & K_{2}^{+}%
\end{array}%
\right] $ in (4.7)$.$ Then%
\begin{eqnarray}
K_{1}^{-}\left( x,y,\tau ;t\right) &=&a_{11}(x,y;t)f_{2}\left( \tau
-x;t\right) f_{1}\left( y+x;t\right) +a_{12}(x,y;t)f_{4}\left( \tau
-x;t\right) f_{1}\left( y+x;t\right) ,  \notag \\
&&  \TCItag{5.7} \\
K_{2}^{-}\left( x,y,\tau ;t\right) &=&a_{11}(x,y;t)f_{3}\left( y+x;t\right)
f_{2}\left( \tau -x;t\right) +a_{22}(x,y;t)f_{3}\left( y+x\right)
f_{4}\left( \tau -x\right) ,  \notag
\end{eqnarray}%
where 
\begin{eqnarray*}
a_{11}(x,y;t) &=&\frac{1-\alpha _{34}\alpha _{43}}{1-\alpha _{12}\alpha
_{21}-\alpha _{34}\alpha _{43}+\alpha _{12}\alpha _{21}\alpha _{34}\alpha
_{43}-\alpha _{34}\alpha _{41}\alpha _{12}\alpha _{23}}, \\
a_{12}(x,y;t) &=&\frac{\alpha _{34}\alpha _{23}}{1-\alpha _{12}\alpha
_{21}-\alpha _{34}\alpha _{43}+\alpha _{12}\alpha _{21}\alpha _{34}\alpha
_{43}-\alpha _{34}\alpha _{41}\alpha _{12}\alpha _{23}}, \\
a_{21}(x,y;t) &=&\frac{\alpha _{12}\alpha _{41}}{1-\alpha _{12}\alpha
_{21}-\alpha _{34}\alpha _{43}+\alpha _{12}\alpha _{21}\alpha _{34}\alpha
_{43}-\alpha _{34}\alpha _{41}\alpha _{12}\alpha _{23}}, \\
a_{22}(x,y;t) &=&\frac{1-\alpha _{12}\alpha _{21}}{1-\alpha _{12}\alpha
_{21}-\alpha _{34}\alpha _{43}+\alpha _{12}\alpha _{21}\alpha _{34}\alpha
_{43}-\alpha _{34}\alpha _{41}\alpha _{12}\alpha _{23}}
\end{eqnarray*}%
with%
\begin{eqnarray*}
\alpha _{21} &=&\int_{y}^{+\infty }f_{2}\left( s-x;t\right) g_{1}\left(
s-x;t\right) ds,\text{ }\alpha _{41}=\int_{y}^{+\infty }f_{4}\left(
s-x;t\right) g_{1}\left( s-x;t\right) ds, \\
\alpha _{23} &=&\int_{y}^{+\infty }f_{2}\left( s-x;t\right) g_{3}\left(
s-x;t\right) ds,\text{ }\alpha _{43}=\int_{y}^{+\infty }f_{4}\left(
s-x;t\right) g_{3}\left( s-x;t\right) ds,
\end{eqnarray*}%
\begin{eqnarray*}
\alpha _{12} &=&\int_{-\infty }^{y}g_{2}\left( s+x\right) f_{1}\left(
s+x\right) ds,\text{ }\alpha _{34}=\int_{-\infty }^{y}g_{4}\left(
s+x;t\right) f_{3}\left( s+x;t\right) ds, \\
\alpha _{12} &=&\int_{-\infty }^{y}g_{2}\left( s+x;t\right) f_{1}\left(
s+x;t\right) ds,\text{ }\alpha _{34}=\int_{-\infty }^{y}g_{4}\left(
s+x;t\right) f_{3}\left( s+x;t\right) ds
\end{eqnarray*}

and%
\begin{eqnarray}
K_{1}^{+}\left( x,y,\tau ;t\right) &=&b_{11}(x,y;t)g_{2}\left( \tau
+x;t\right) g_{1}\left( y-x;t\right) +b_{12}(x,y;t)g_{2}\left( \tau
+x;t\right) g_{3}\left( y-x;t\right) ,  \notag \\
&&  \TCItag{5.8} \\
K_{2}^{+}\left( x,y,\tau ;t\right) &=&b_{21}(x,y;t)g_{4}\left( \tau
+x;t\right) g_{1}\left( y-x;t\right) +b_{22}(x,y;t)g_{4}\left( \tau
+x;t\right) g_{3}\left( y-x;t\right) ,  \notag
\end{eqnarray}%
where%
\begin{eqnarray*}
b_{11}(x,y;t) &=&\frac{1-\beta _{34}\beta _{43}+\beta _{12}\beta _{23}\beta
_{34}\beta _{41}}{1-\beta _{12}\beta _{21}-\beta _{34}\beta _{43}+\beta
_{12}\beta _{21}\beta _{34}\beta _{43}}, \\
b_{12}(x,y;t) &=&\frac{\beta _{34}\beta _{41}}{1-\beta _{12}\beta
_{21}-\beta _{34}\beta _{43}+\beta _{12}\beta _{21}\beta _{34}\beta _{43}},
\\
b_{21}(x,y;t) &=&\frac{\beta _{12}\beta _{23}}{1-\beta _{12}\beta
_{21}-\beta _{34}\beta _{43}+\beta _{12}\beta _{21}\beta _{34}\beta _{43}},
\\
b_{22}(x,y;t) &=&\frac{1-\beta _{12}\beta _{21}+\beta _{12}\beta _{23}\beta
_{34}\beta _{41}}{1-\beta _{12}\beta _{21}-\beta _{34}\beta _{43}+\beta
_{12}\beta _{21}\beta _{34}\beta _{43}}
\end{eqnarray*}%
with%
\begin{eqnarray*}
\beta _{12} &=&\int_{-\infty }^{y}g_{2}\left( s+x;t\right) f_{1}\left(
s+x;t\right) ds,\text{ }\beta _{32}=\int_{-\infty }^{y}g_{2}\left(
s+x;t\right) f_{3}\left( s+x;t\right) ds, \\
\beta _{14} &=&\int_{-\infty }^{y}g_{4}\left( s+x;t\right) f_{1}\left(
s+x;t\right) ds,\text{ }\beta _{34}=\int_{-\infty }^{y}g_{4}\left(
s+x;t\right) f_{3}\left( s+x;t\right) ds, \\
\beta _{21} &=&\int_{y}^{+\infty }f_{2}\left( s-x;t\right) g_{1}\left(
s-x;t\right) ds,\text{ }\beta _{41}=\int_{y}^{+\infty }f_{4}\left(
s-x;t\right) g_{1}\left( s-x;t\right) ds, \\
\beta _{23} &=&\int_{y}^{+\infty }f_{2}\left( s-x;t\right) g_{3}\left(
s-x;t\right) ds,\text{ }\beta _{43}=\int_{y}^{+\infty
}f_{4}(s-x;t)g_{3}\left( s-x;t\right) ds.
\end{eqnarray*}

Thus, the explicit solution of the equation (5.2) with the potential (5.1)
having the scattering data in the degenerate form of (5.6), exists and it is
in the following form:\textit{\ }%
\begin{eqnarray*}
q_{1}(x,y,y;t) &=&-2K_{1}^{-}\left( x,y,y;t\right) ,\text{ }%
q_{2}(x,y,y;t)=-2K_{2}^{-}\left( x,y,y;t\right) , \\
q_{3}(x,y,y;t) &=&-2K_{1}^{+}\left( x,y,y;t\right) ,\text{ }%
q_{4}(x,y,y;t)=-2K_{2}^{+}\left( x,y,y;t\right) ,
\end{eqnarray*}%
where\textit{\ }$K_{1}^{-}\left( x,y,\tau ;t\right) $\textit{\ }and\textit{\ 
}$K_{2}^{-}\left( x,y,\tau ;t\right) $\textit{\ }are determined by (5.7),%
\textit{\ }$K_{1}^{+}\left( x,y,\tau ;t\right) $\textit{\ }and\textit{\ }$%
K_{2}^{+}\left( x,y,\tau ;t\right) $\textit{\ }are determined by (5.8)%
\textit{. }

\section{\textbf{Conclusion}}

In this paper, we consider the inverse scattering method for a
generalization of both systems, the Davey-Stewartson system in two spatial
dimensions\ and 2n-wave interaction problem with $n$ velocities that
examines the Benney-type model of interactions between short and long waves.
We show the existence of a unique solution for Cauchy problem on an
arbitrary interval of time for small initial data and also the
multi-solitons corresponding to degenerate kernels of GLM-type equation that
is associated with solutions of the two dimensional analogue of Manakov
system. The construction of multi-solitons within inverse scattering
techniques for a more generalized two dimensional nonlinear evolutional wave
equations will be considered as a future work.


\begin{thebibliography}{99}
\bibitem{[1]} R. Beals and R. R. Coifman, \textit{Linear spectral problem,
nonlinear equations and the }$\partial $\textit{-bar method}, Inverse
Problems 5 (1989), 87--130.

\bibitem{[2]} H. Cornille, \textit{Solutions of the generalized nonlinear
Schr\"{o}dinger equation in two spatial dimensions}, J. Math. Phys. 20
(1978), no. 1, 199-209.

\bibitem{[3]} Fokas A S and Ablowitz M J., \textit{On the inverse scattering
of the time dependent Schr\"{o}dinger equation and the associated KPI
equation,} Stud. Appl. Math. 69 (1983), 211--28.

\bibitem{[4]} A. S. Fokas and M. J. Ablowitz, \textit{Methods of solution
for a class of multi-dimensional nonlinear evolution equations,} Phys. Rev.
Lett. 51 (1983), 7--10.

\bibitem{[5]} A. S. Fokas and L. Y. Sung, \textit{On the solvability of the }%
$N$\textit{- Wave, Davey-Stewartson and Kodomtsev-Petviashvili equations,}\
Inverse Problems \textbf{8 (}1992\textbf{)}, 673-708.

\bibitem{[6]} A. S. Fokas and M. J. Ablowits, \textit{On the inverse
scattering transform of multidimensional nonlinear equations related to
first-order systems in the plane}, J. Math. Phys. \textbf{25 (}1984\textbf{) 
}no. 8, 2494-2505.

\bibitem{[7]} N. Sh. Iskenderov and M. I. Ismailov, \textit{On the inverse
scattering transform of a nonlinear evolution equation with 2+1 dimensions
related to nonstrict hyperbolic systems}. Nonlinearity 25 (2012), no. 7,
1967--1979.

\bibitem{[8]} M. I. Ismailov, \textit{Inverse scattering problem for
nonstationary Dirac-type systems on the plane,} Journal of Mathematical
Analysis and Applications, \textbf{365 (}2010), 498-509.

\bibitem{[9]} M. I. Ismailov,\textit{\ Integration of nonlinear system of
4-waves with two velocities in 2 + 1 dimensions by the inverse scattering
transform method, }Journal of Mathematical Physics 52 (2011), 033504.

\bibitem{[10]} S. V. Manakov, \textit{On the theory of two-dimensional
stationary self-focusing of electromagnetig waves},\ Sov. Phys. JETP 38
(1974), 248-253.

\bibitem{[11]} L. P. Nizhnik, \textit{Integration of multidimensional
nonlinear equations by the inverse problem method.} Soviet Phys. Dokl. 25
(1980), no. 9, 706--708.

\bibitem{[12]} L. . Nizhnik, M. D. Pochinaiko, \textit{Integration of a
spatially two-dimensional nonlinear Schr\"{o}dingerequation by the inverse
problem method}, Functional Anal. Appl. 16 (1982), no. 1, 66--69.

\bibitem{[13]} L. P. Nizhnik, \textit{The inverse scattering problem for
hyperbolic equations and their application to nonlinear integrable systems.}
Reports on Math \ Phys. \textbf{26} (1988), no.2, 261-283.

\bibitem{[14]} L. P. Nizhnik, \textit{An inverse problem of nonstationary
scattering for the Dirac equations,} Ukr. Mat. Zh., 24 (1972), no. 1,
112-115.

\bibitem{[15]} R. G. Novikov, \textit{Inverse scattering up to smooth
functions for the Dirac-ZS-AKNS system}, Selecta Math. (N.S.) 3 (1997), no.
2, 245--302.

\bibitem{[16]} A. L. Sakhnovich, \textit{Dirac type system on the axis:
explicit formulae for matrix potentials with singularities and
soliton-positon interactions}, Inverse Problems 19 (2003), no. 4, 845--854.

\bibitem{[17]} L. Y. Sung, and A. S. Fokas, \textit{Inverse problem for }$%
N\times N$\textit{\ hyperbolic systems on the plane and the }$N-$\textit{%
wave interactions}. Comm. Pure Appl. Math. \textbf{64 (}1991\textbf{)},
535-571.

\bibitem{[18]} V. E. Zakharov, A. B. Shabat, \textit{The scheme of
integration of nonlinear equations of mathematical physics by inverse
scattering method. I,} Funct. Anal. Appl. 8 (1974), no. 3, 226-235;

\bibitem{[19]} V. E. Zakharov, A. B. Shabat, \textit{Integration of
nonlinear equations of mathematical physics by the method of inverse
scattering}. II Funct. Anal. Appl. 13 (1979), no. 3, 166-174.

\bibitem{[20]} D. J. Benney, \textit{A general theory for interactions
between short and long waves}, Stud. Appl. Math.. 56 (1977), 81 - 94.
\end{thebibliography}
\end{document}